\documentclass[twocolumn,
notitlepage,
superscriptaddress,
nofootinbib,
 bibnotes,
 amsmath,amssymb,
 aps,
 prl,
]{revtex4-2}

\usepackage{amsmath}
\usepackage{graphicx}
\usepackage{bm}
\usepackage{physics}
\usepackage{color}
\usepackage{soul}
\usepackage{appendix}
\usepackage{multirow}
\usepackage{float}
\restylefloat{table}
\usepackage{mathtools}
\usepackage{comment}
\usepackage{textcomp}
\usepackage[justification=raggedright,singlelinecheck=false]{caption}
\usepackage{subcaption}
\usepackage{amsthm}
\usepackage{dsfont} %Helps me get the right symbol for identity
\usepackage[colorlinks=true, urlcolor=blue]{hyperref}
\usepackage{MnSymbol}

\usepackage{tikz}
\usepackage[export]{adjustbox}

\newcommand{\fket}[1]{\left.\middle|\,#1\right)}

\newcommand{\fbraket}[2]{\left( #1\,\middle|\,  #2 \right)}

\begin{document}

\title{Diameter truncated operator evolution}

\author{Tom Holden-Dye}
\email[Corresponding author: ]{thomas.holden-dye.21@ucl.ac.uk }
\affiliation{Department of Physics and Astronomy, University College London, Gower Street, London, WC1E 6BT, UK}
\author{Max Marvell}
\affiliation{Department of Physics, King's College London, Strand WC2R 2LS, UK}
\author{Joel Mills}
\affiliation{Phasecraft Ltd, London, UK}
\affiliation{Department of Physics and Astronomy, University College London, Gower Street, London, WC1E 6BT, UK}
\author{Christopher J. Turner}
\affiliation{Department of Physics and Astronomy, University College London, Gower Street, London, WC1E 6BT, UK}
\author{Arijeet Pal}
\affiliation{Department of Physics and Astronomy, University College London, Gower Street, London, WC1E 6BT, UK}

\begin{abstract}
    We present a method for simulating operator dynamics in out-of-equilibrium quantum systems. Due to the rapid growth of complexity in these systems, this is typically speaking an intractable task. However, exceptional progress has been made in recent years to sidestep this barrier, with the introduction of a number of methods that make use of a truncation of the simulation to low-\emph{weight} (the number of non-trivial terms in a Pauli string basis expansion) observables, which turns out to be a good approximation for many dynamical quantities of interest, e.g., two-point infinite-temperature correlation functions between local operators. In this work, we extend this idea to a leaner truncation protocol, truncating operators based on their \emph{diameter} - that is, the size of the region on the lattice on which they are non-trivially supported. Using existing analysis for generic circuits we argue that this kind of truncation protocol is physically well-motivated, and show via extensive numerical simulations for a number of systems of interest (here, the kicked Ising model and the Heisenberg XXZ model) that it is effective, and allows us to efficiently and accurately extract local correlation functions and transport properties.
\end{abstract}

\maketitle
The vast majority of many-body quantum systems thermalise. Left to evolve from some initial state, they will eventually relax - at least from the point of view of local probes - to equilibrium \cite{deutsch1991ETH,srednicki1994ETH,rigol2008ETH,gogolin2016ETHreview}. In this late-time limit (i.e. once local thermalisation has been achieved), a simple description emerges: finite-sized subsystems are described by an appropriate Gibbs state \cite{popescu2006entanglementandstatmech,goldstein2006canonicaltypicality,rigol2007relaxationintegrablembs}, that depends only on a small handful of ``slow" degrees of freedom (typically tied to conserved quantities), the relaxation of which can be described by a coarse-grained, hydrodynamical theory \cite{kadanoff1963hydrodynamics,doyon2020GHDlecturenotes}. This approach has proven extraordinarily successful in predicting universal, late-time dynamical behaviour \cite{lux2014hydrodynamiclongtimetails,bertini2021finiteTtransport1D}, including deviations from diffusion (e.g. subdiffusion in kinetically-constrained models \cite{gromov2020fractonhydrodynamics}, and superdiffusion in the 1D Heisenberg XXX chain \cite{gopalakrishnan2019kinetictheoryXXZ}).\footnote{We have referred at points here to work concerning \emph{integrable} models \cite{Korepin1993}. We study them herein (alongside non-integrable models) as they represent examples of models with non-trivial transport phenomena with robust theoretical predictions for us to compare numerics against. We note, however, that these models  do not thermalise in the conventional sense (though they do still equilibrate to generalised  Gibbs ensembles) \cite{caux2013timeevolutionquenchintegrable, Vidmar_2016}.}
\par
In general, however, it is much more challenging to study the \emph{approach} to equilibrium - to study the behaviour of \emph{thermalising} systems at short or intermediate times. Ultimately, this is because they quickly become very complex: the entanglement of out-of-equilibrium states grows rapidly \cite{calabrese2005entanglemententropy1D,kim2013ballisticspreadingentanglementdiffusive}; so too do other quantum resources, such as nonstabiliserness and non-Gaussianity \cite{turkeshi2025magicspreading,zhang2026magicdynamics,tirrito2025anticoncentrationmagicspreading,aditya2025growthspreadingquantumresources,montanalopez2024exactsolutionlongrangesre}; and initially local operators fast become highly non-local and ``scrambled" \cite{nahum2018opspreadingRUCs,vonkeyserlingk2018ophydrodynamicsRUCs}. This fundamentally hampers the accuracy of traditional tensor-network based simulations beyond very short times; it is impossible to accurately capture the dynamical evolution of the full many-body wavefunction, or of all the possible observables, efficiently \cite{prosen2007efficiencyclassicalsimulations,schuch2008onentropygrowthhardnesssimulating,dowling2025magicresourcesheisenbergpicture,dowling2026classicalsimulabilityoperatorentanglement}.

Despite this, significant progress has been made in recent years in addressing this challenge, with the introduction of a number of numerical methods that are capable of extracting hydrodynamical quantities, such as transport coefficients, from first principles calculations in strongly-interacting lattice systems. Central to these methods is an important observation: while operators tend to scramble and become highly non-local, often the majority of the operator weight generated does not contribute to dynamical quantities of interest \cite{vonkeyserlingk2022operatorbackflow}. In particular, overlaps with other local operators after some time (two-point correlation functions) can often be well approximated by retaining only a small subset of relevant operator trajectories \cite{rakovszky2022DAOE}.

\begin{figure*}
    \includegraphics[width=\textwidth]{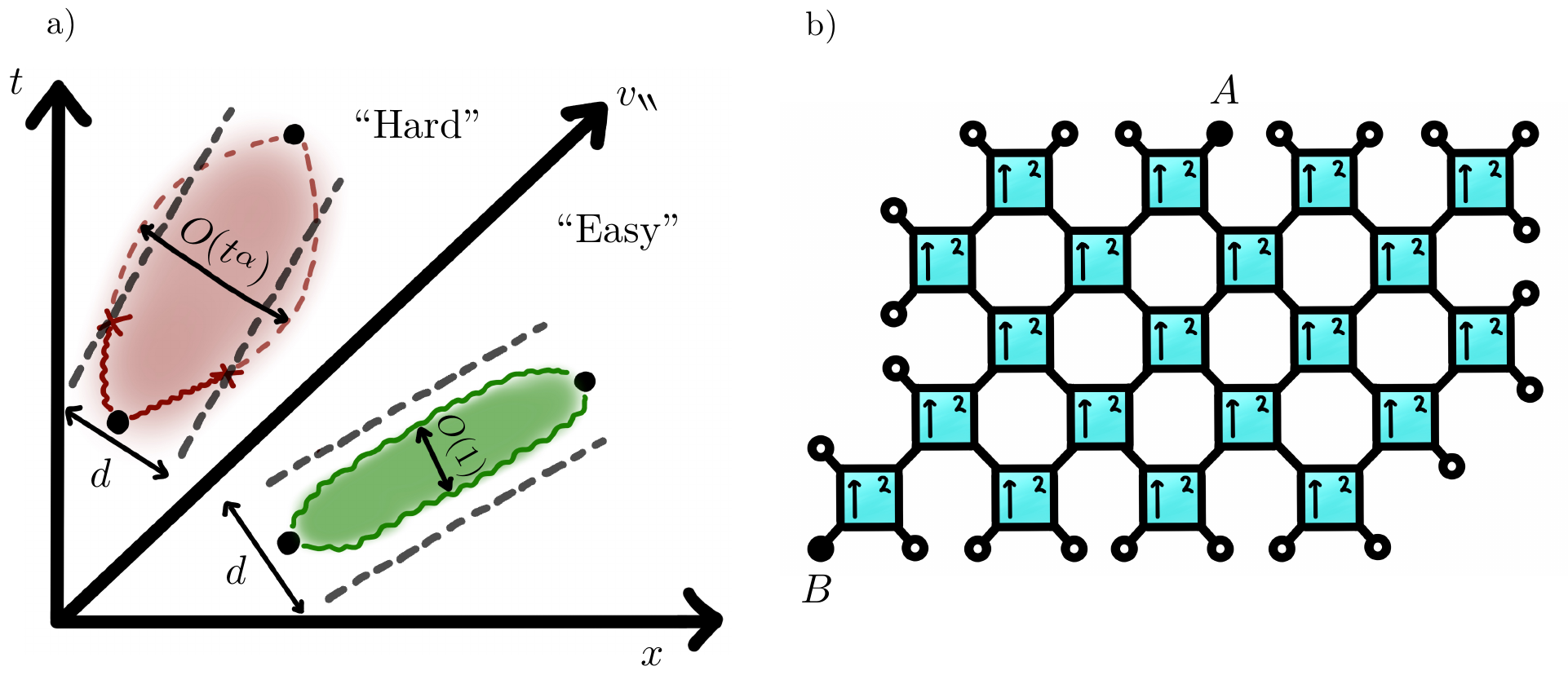}
    \caption{Diameter-truncated operator evolution (DTOE). a) The method works by truncating out operators that extend in real-space beyond some threshold diameter $d$. In a wide-range of lattice systems, there is an extensive region - typically above some critical velocity, $v_c$ - within which two-point correlation functions of local observables are dominated by trajectories through operators of a bounded, constant diameter \cite{nahum2022realtimecorrelatorsinchaoticmbs}. This is then an ``easy" phase for DTOE, in which it can approximate two-point functions with $O(t)$ memory. In the rest of the lattice, trajectories through operators with $O(t^{\alpha})$ diameter (with $\alpha \leq 1$) are expected to dominate, leading to a (sub-)exponential memory scaling. b) Two-point functions $\fbraket{A_x}{B_0(t)}$ are computed by chaining together transfer matrices that map intervals of diameter-$d$ operators left and right on the chain, generating a path between the two operators. This generalises path-sum approaches developed for perturbed dual-unitary circuits \cite{kos2021ducpathint} to arbitrary width paths.}\label{fig:schematic}
\end{figure*}

The challenge, then, is to ascertain how to consistently restrict a simulation of the system to only the relevant operator trajectories. The methods of dissipation-assisted operator evolution (DAOE) \cite{rakovszky2022DAOE} and Pauli Propagation \cite{rudolph2026paulipropagationcomputationalframework} do so by truncating out high-\emph{weight} operator content: that is, components of the time-evolving operator, when expanded in an operator (Pauli) string basis, that have a high number of non-trivial (non-identity) terms. While such an approximation necessarily chops out a lot of operator weight, it keeps the important, low-complexity content that contributes to the local observables (e.g. two-point functions of local operators) one usually cares about.

In this work, we introduce \emph{diameter-truncated operator evolution} (DTOE). This is a closely related method to the two mentioned above, but differs in that instead of truncating operators based on their weight, operators are truncated based on their \emph{diameter} - that is, the real-space size of the region of the lattice on which the operator is non-trivially supported. We argue that for a wide range of systems, this truncation is well motivated - it keeps precisely the most relevant content necessary for computing two-point functions. In addition, it is in principle more efficient than weight based methods - on a 1D lattice of $L$ $q$-dimensional qudits, there are $N_{\textrm{weight}-d} = {L \choose d} (q^2-1)^d = O(L^d)$ operators of weight $d$, whereas there are $N_{\textrm{diameter}-d} = L(q^2-1)q^{2(d-1)} = O(L)$ operators of diameter $d$, and so $N_{\textrm{diameter}-d}/N_{\textrm{weight}-d} \rightarrow 0$ in the limit of large $L$ for all $d \geq 2$.
\par
We demonstrate the efficacy of the method with simulations of integrable and non-integrable systems. We report accurate computations of infinite-temperature two-point functions in both cases at time-scales beyond what has been achieved (to the best of our knowledge) in simulations of comparable models performed by current quantum computers \cite{google2024dynamicsofmagnetization, fischerIBM2026dynamicalsimulationsofmbqc}. No high-performance compute was used in the generation of our results. We present everything here in the framework of 1+1D brickwork quantum circuits, and with a hard truncation scheme (all operators above a certain cut-off diameter are strictly omitted from the simulation), but comment on extensions to more general settings (higher lattice dimensions) and softer truncation schemes.

\noindent \textit{Diameter-truncated operator evolution}---We start by presenting the method in a simple form. We will be considering 1+1D brickwork quantum circuits, i.e., some unitary
\begin{equation}
    \mathcal{U}(t) = \prod_{t^{\prime}=1}^t U_{t^{\prime}} \in U(q^{2L})
\end{equation}
that acts on a 1D lattice $\Lambda = \mathbb{Z}_{2L}$ of $2L$ $q$-dimensional qudits. We will denote a basis of (generalised) Pauli operators on each qudit as $\{\sigma^{(i)}\}_{i=0}^{q^2-1}$, with $\sigma^{(0)} = \mathds{1}$. Each unitary $U_{t^{\prime}}$ is a layer of neighbouring 2-qudit gates $\texttt{u} = \includegraphics[width=0.065\linewidth, valign=c]{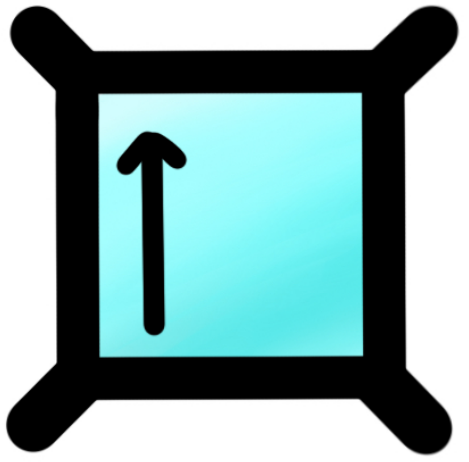} \in U(q^2)$, and these are staggered in an odd-even, ``brickwork" pattern:
\begin{equation}
\mathcal{U}(t) =
\begin{tikzpicture}[baseline=(img.center)]
  % the circuit image, anchored so its lower-left corner is the origin
  \node[anchor=south west,inner sep=0] (img) at (0,0)
    {\includegraphics[width=0.23\textwidth]{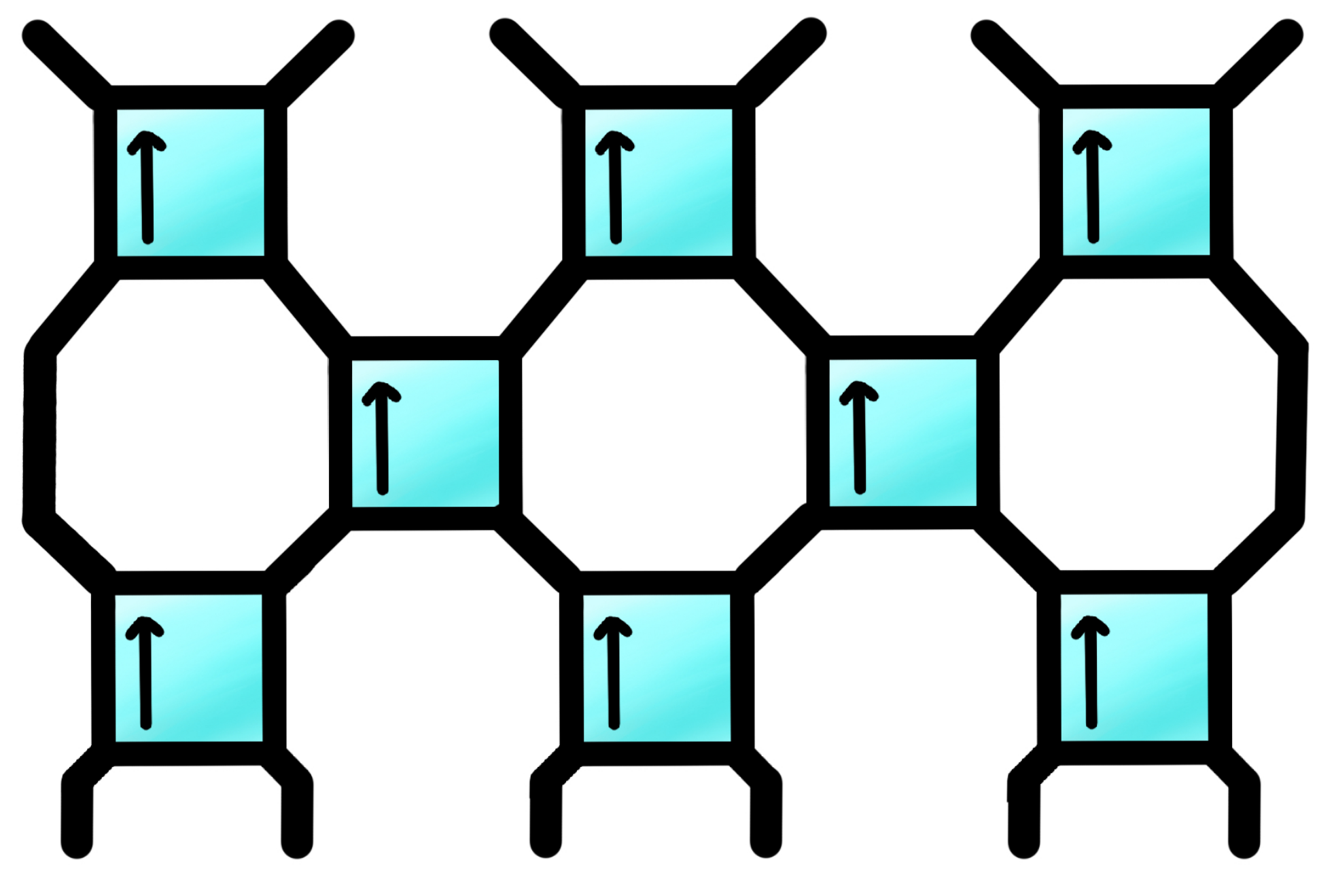}};
  % normalised overlay: (0,0)=bottom-left of image, (1,1)=top-right
  \begin{scope}[x={(img.south east)},y={(img.north west)}]
    % --- time gridlines (dashed) extending to the right, with t-labels ---
    % y-fractions tuned to the three layers of gates + the t=0 leg row
    \def\tzero{0.075}   % bottom legs  (t=0)
    \def\tone{0.39}     % middle gate row (t=1)
    \def\ttwo{0.65}     % top gate row (t=2)
    \foreach \yy/\lab in {\tzero/{t=0},\tone/{t=1},\ttwo/{t=2}}{
      \draw[dashed] (0,\yy) -- (1.08,\yy);
      \node[anchor=west] at (1.09,\yy) {$\lab$};
    }
    % vertical dots above the top row
    \node[anchor=south] at (1.21,0.75) {$\vdots$};
    % --- x-axis site labels along the bottom ---
    % fractions tuned to the three columns of bottom legs
    %\node[anchor=north] at (-0.2,-0.04)  {$x =$};
    \node[anchor=north] at (0.01, 0)  {$-L+1$};
    \node[anchor=north] at (0.29,-0.045) {$\cdots$};
    \node[anchor=north] at (0.415,0)  {$0$};
    \node[anchor=north] at (0.58,0) {$1$};
    \node[anchor=north] at (0.7,-0.045) {$\cdots$};
    \node[anchor=north] at (0.94,0)  {$L$};
  \end{scope}
\end{tikzpicture}
\label{eqn:brickwork}
\end{equation}
We will take $L = t$ to avoid finite-size effects. For brevity, we assume here that the circuit is homogeneous (the same 2-site unitary is used everywhere), but stress that the approach to be described immediately generalises (with a bit of bookkeeping) to the inhomogeneous case.
\par
We will be interested in computing infinite-temperature two-point correlation functions
\begin{equation}
    C_{AB}(x,t) = \tr(A_x^{\dagger}B_0(t)\rho_{\infty}),
\end{equation}
where $A$ and $B$ are local operators and $\rho_{\infty} = (\mathds{1}/q)^{\otimes 2L}$ is the infinite-temperature state. We will do this, however, in a restricted space of operators,
\begin{equation}
    V_d = \bigoplus_{x \in \Lambda} \left[\mathcal{P}_d\right]_x,
\end{equation}
where to each site $x \in \Lambda$ we have assigned the set of $d$-letter Pauli words with the left-most\footnote{Or right-most - it doesn't matter, as long as one is consistent.} Pauli restricted to be non-identity:
\begin{equation}
    \mathcal{P}_d = \{P^{(\vec{s})} = \bigotimes_{i \in \mathbb{Z}_d} \sigma^{(s_i)} \; | \; \vec{s} \in \mathbb{Z}_{q^2}^d, \; s_0 \neq 0 \}.
\end{equation}
We will then be interested in constructing the dynamics of the circuit restricted to this space. We first briefly introduce folded notation: we define an operator-to-state vectorisation mapping,
\begin{equation}
    \ket{a}\bra{b} \mapsto \ket{a} \otimes \ket{b}^*, 
\end{equation}
for computational basis (or any other suitable basis for $\mathbb{C}^q$) states $\ket{a}$ and $\ket{b}$ \cite{bertini2020opent1}. This maps operators on $\mathbb{C}^q$ to vectors on a doubled Hilbert space $\mathbb{C}^q \otimes \mathbb{C}^q$. We will denote the vectorisation of our initial operator $B_0$ as $\fket{B_0}$. Denoting complex conjugation with a darker shading of the gate, $\texttt{u}^* = \includegraphics[width=0.065\linewidth, valign=c]{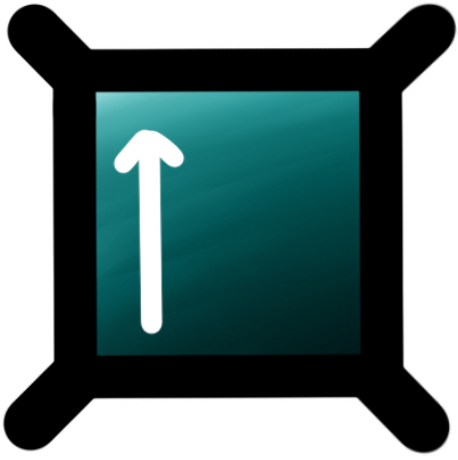}$, we then define ``folded" gates $\texttt{v} \in U(q^4)$,
\begin{equation}
    \texttt{v} = \includegraphics[width=0.09\linewidth, valign=c]{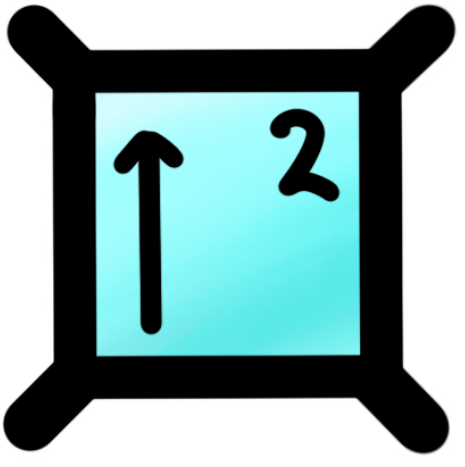} = \includegraphics[width=0.1\linewidth, valign=c]{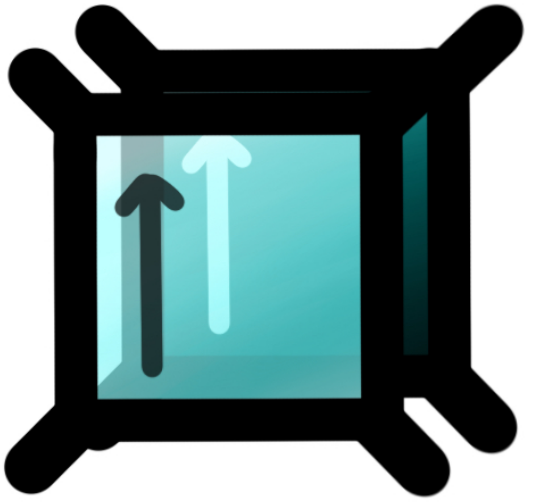} = \includegraphics[width=0.09\linewidth, valign=c]{U.pdf} \otimes \includegraphics[width=0.09\linewidth, valign=c]{Ustar.pdf} = \texttt{u} \otimes \texttt{u}^*,
\end{equation}
that act on vectors in the doubled Hilbert space, representing the adjoint action $\texttt{u}A\texttt{u}^{\dagger}$ of $\texttt{u}$ on operators $A$ on the original unfolded Hilbert space. Finally, we denote the vectorisation of the identity as
\begin{equation}
    \fket{\medcircle} = \;\; \includegraphics[width=0.04\linewidth, valign=c]{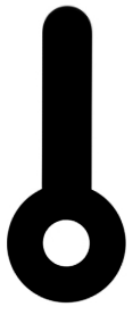} = \;\; \includegraphics[width=0.05\linewidth, valign=c]{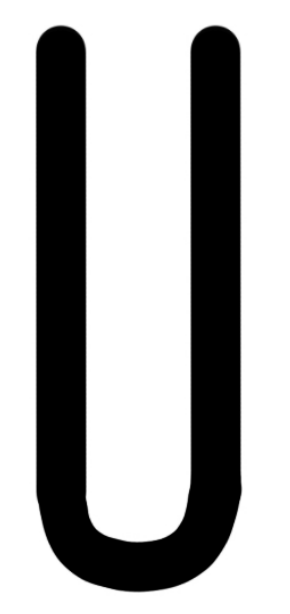} = \sum_{i=0}^{q-1} \ket{ii},
\end{equation}
 Using this notation, we then define the following maps:
\begin{equation}
    \mathcal{M}_d^{0,0} := \begin{cases}
        \includegraphics[width = 0.07\linewidth, valign = c]{U2.pdf}^{\otimes \frac{d-1}{2}} \otimes \includegraphics[width = 0.08\linewidth, valign = c]{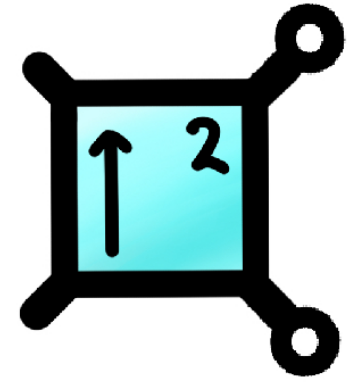}\;, &\textrm{$d$ odd,} \\
        \includegraphics[width = 0.07\linewidth, valign = c]{U2.pdf}^{\otimes \frac{d}{2}}, &\textrm{$d$ even,} 
    \end{cases}
\end{equation}
\begin{equation}
    \mathcal{M}_d^{0,1} := \begin{cases}
        \includegraphics[width = 0.08\linewidth, valign = c]{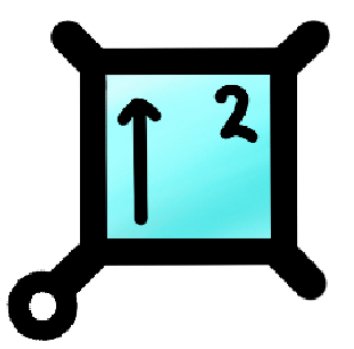} \otimes \includegraphics[width = 0.07\linewidth, valign = c]{U2.pdf}^{\otimes \frac{d-3}{2}} \otimes \includegraphics[width = 0.08\linewidth, valign = c]{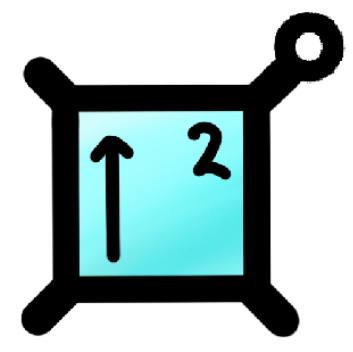}\;, &\textrm{$d$ odd,} \\
        \includegraphics[width = 0.08\linewidth, valign = c]{U0010.pdf} \otimes \includegraphics[width = 0.07\linewidth, valign = c]{U2.pdf}^{\otimes \frac{d}{2}-1} \otimes \includegraphics[width = 0.08\linewidth, valign = c]{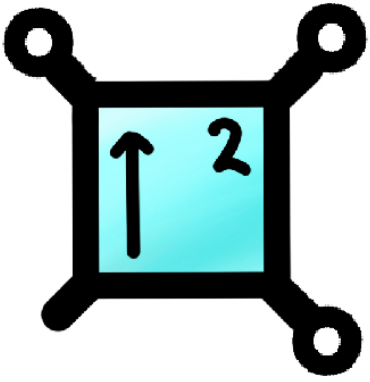}\;, &\textrm{$d$ even.}
    \end{cases}
\end{equation}
\begin{equation}
    \mathcal{M}_d^{1,0} := \begin{cases}
        \includegraphics[width = 0.08\linewidth, valign = c]{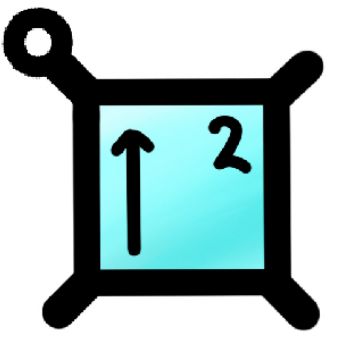} \otimes \includegraphics[width = 0.07\linewidth, valign = c]{U2.pdf}^{\otimes \frac{d-3}{2}} \otimes \includegraphics[width = 0.08\linewidth, valign = c]{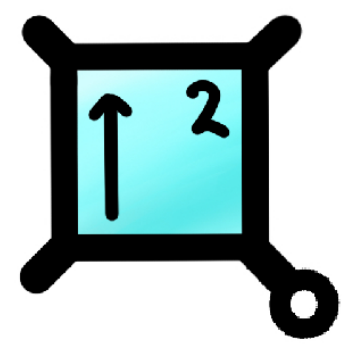}\;, &\textrm{$d$ odd,} \\
        \includegraphics[width = 0.08\linewidth, valign = c]{U1000.pdf} \otimes \includegraphics[width = 0.07\linewidth, valign = c]{U2.pdf}^{\otimes \frac{d}{2}-1} \otimes \;\; \includegraphics[width = 0.02 \linewidth, valign = c]{Circ.pdf}\;, &\textrm{$d$ even,}
    \end{cases}
\end{equation}
\begin{equation}
    \mathcal{M}_d^{1,1} := \begin{cases}
        \includegraphics[width = 0.08\linewidth, valign = c]{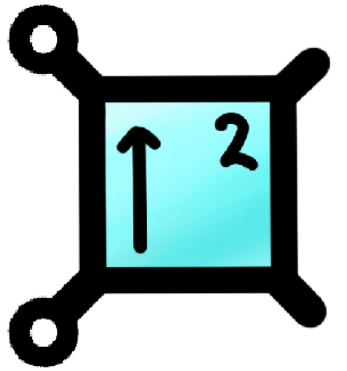} \otimes \includegraphics[width = 0.07\linewidth, valign = c]{U2.pdf}^{\otimes \frac{d-1}{2}}, &\textrm{$d$ odd,} \\
        \includegraphics[width = 0.08\linewidth, valign = c]{U1010.pdf} \otimes \includegraphics[width = 0.07\linewidth, valign = c]{U2.pdf}^{\otimes \frac{d}{2}-1} \otimes \includegraphics[width = 0.08\linewidth, valign = c]{U0101.pdf}\;, &\textrm{$d$ even,}
    \end{cases}
\end{equation}
We then use these maps to define the map
\begin{equation}
    \mathcal{M}_{d} = \begin{pmatrix}
        \mathcal{M}_d^{0,0} & \mathcal{M}_d^{0,1} \\
        \mathcal{M}_d^{1,0} & \mathcal{M}_d^{1,1} \\
    \end{pmatrix}.
\end{equation}
which we use to update our initial state
\begin{equation}
    \fket{B_0(t)}_d = \prod_{t^{\prime} = 1}^t \bigoplus_{x: x \oplus t^{\prime} = 1} \left[\mathcal{M}_{d}\right]_{x,x+1}\fket{B_0}.
\end{equation}
The correlation functions are then given as
\begin{equation}
    C_{AB}(x,t)_d = \fbraket{A_x}{B_0(t)}_d.
\end{equation}
Normalisation can be taken care of by performing all calculations in the Pauli basis, and taking the operators $A$ and $B$ to be orthonormal in this basis.

\noindent \textit{Physical motivation}---The method we have described above computes approximations $C_{AB}(x,t)_d$ to the true two-point functions $C_{AB}(x,t)$ by restricting the dynamics at all time to the space of diameter-$d$ operators. Assuming we keep $d = O(1)$ constant, the time-complexity of this method should scale loosely as $O(t^2)$. This is very efficient, but we are of course making a very big approximation - the dynamics on $V_d$, is, generally speaking, vastly different to the full dynamics on $\mathcal{H} = \left(\mathbb{C}^{q}\right)^{\otimes 2L}$ (for one we expect $||A_0(t)||_{V_d}/||A_0(t)||_{\mathcal{H}} \rightarrow 0$ as $t \rightarrow \infty$). Why should we expect this to be a reasonable approximation?
\par
The key point is that while the majority of the operator norm follows trajectories into the space of highly non-local operators, local two-point correlation functions are, for the most part, insensitive to these trajectories - it is the \emph{atypical} trajectories, which return to the space of local operators, that contribute. Ref.~\cite{nahum2022realtimecorrelatorsinchaoticmbs} studied the form of these trajectories - those that dominate the two-point functions - in chaotic quantum systems. Crucially, they found that for most systems, there exists at least some region of the space-time lattice - typically the section above some critical velocity $v_c$ - for which the dominant trajectories are those that travel exclusively through the space of operators with a constant $O(1)$-sized region of support (i.e. through $V_d$ for some fixed, time-independent $d$). Outside of this region, the dominant trajectories will travel through operators of $O(t^{\alpha})$ size, for some $\alpha > 0$.
\par
These two regions represent two different phases, in terms of the complexity of approximating two-point functions within them with DTOE: there is the ``easy" phase - the $O(1)$-region - in which two-point functions can be faithfully approximated with a constant cut-off diameter-$d$ (leading to a polynomial time-complexity for DTOE); and there is a ``hard" phase - the $O(t^{\alpha})$ region - in which the cut-off diameter must grow with time in order to faithfully approximate the two-point functions (leading to at least a subexponential time-complexity for DTOE).
\par
For 1+1D Haar-random circuits, it was shown that the critical velocity $v_c$ is equal to the butterfly velocity $v_B = (q^2-1)/(q^2+1)$ ($v_c = v_B = 3/5$ for qubits, $q=2$) \cite{nahum2022realtimecorrelatorsinchaoticmbs}. This means the majority of the space-time lattice - the region $[-v_ct, v_ct]$ - is in the ``hard" phase.\footnote{Note that brickwork circuits have a maximum speed of information propagation (an effective speed of light) of $v = 1$, so all (non-zero) correlations must be contained in the region $[-t+1, t]$; for all $q\geq 2$, we have that $|[-v_ct, v_ct]| \geq 2|[(1-v_c)t, t]|$, and so the hard region is larger than the easy region, when we exclude the trivial region outside of the causal light cone ($|x| > t$).}
\par
Away from Haar-random circuits, however, and away from 1D, the size of the easy region can be much larger. In Ref.~\cite{nahum2022realtimecorrelatorsinchaoticmbs}, the authors note that it is possible to construct circuits in which the whole space-time lattice is in the easy phase. A noteworthy example of this are \emph{perturbed dual-unitary circuits} \cite{kos2021ducpathint} - we present results on the efficacy of DTOE for studying these circuits below.
\par
The presence of conserved charges will also affect the nature of the dominant trajectories to two-point functions, and hence the ``easy"-``hard" split of the space-time lattice \cite{nahum2022realtimecorrelatorsinchaoticmbs}. For generic conserved charge dynamics, one expects the relevant correlation functions (those corresponding to where the charge is concentrated) to be dominated by ``thin" ($O(1$)-support) trajectories through local operators associated to the conserved density \cite{rakovszky2018diffusive, khemani2018conslawsRUCs}. We test this below in the Heisenberg XXZ model, which has a $U(1)$ total-$S_z$ conservation law for all coupling strengths. We extract hydrodynamic scaling exponents for the charge transport using DTOE, and show remarkable agreement with theoretical predictions, even at the isotropic point (where increasingly long strings contribute significantly to the transport with increasing $t$~\cite{gopalakrishnan2019kinetictheoryXXZ}, which might be expected to make things harder for DTOE).
\par
Intriguingly, the authors of Ref.~\cite{nahum2022realtimecorrelatorsinchaoticmbs} argue that in 2D, the whole space-time lattice is much more generically speaking (even in the Haar-random case) expected to be in the ``easy" phase. This suggests DTOE may work surprisingly well in higher lattice dimensions - we do not, however, attempt such a generalisation here.

\begin{figure*}
    \includegraphics[width=\textwidth]{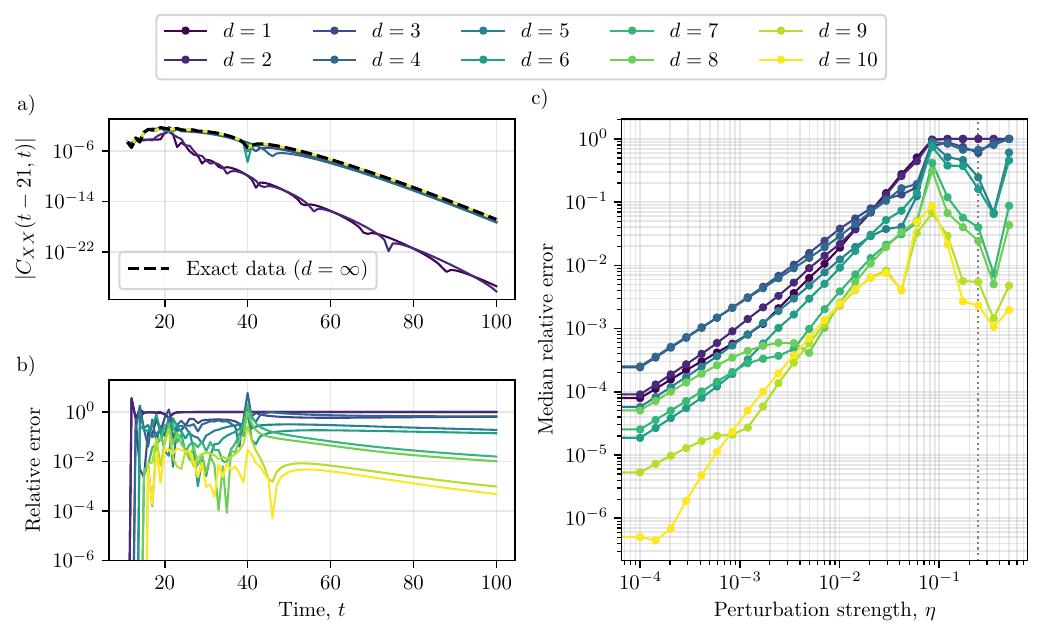}
    \caption{
        DTOE on the kicked Ising model, $U_{KI}[J=B=\pi/4 + \eta, h = 1.2]$. a) The magnitude of $C_{XX}(x,t)_d$, the diameter-$d$ approximation to $C_{XX}(x,t)_d$ (the dashed line), against $t$ with $x = t - 21$, for $d \in [1,10]$, at $\eta = 0.24588$. b) The relative error $(|C_{XX}(x,t)| - |C_{XX}(x,t)_d|)/|C_{XX}(x,t)|$ against $t$ for each diameter at $\eta = 0.24588$. c) The median (over the $t \in [1,100]$ window) relative error in the $x_{-} = 21$ correlators for each diameter over the full range of $\eta \in [0,0.5]$ studied.}\label{fig:pduc}
\end{figure*}

\noindent \textit{Perturbed dual-unitary circuits}---Dual-unitaries are two-site unitary gates which are additionally unitary in the spatial direction of the circuit \cite{bertini2019exactcorrfuncs,gopalakrishnan2019finitedepthinfinitewidth,bertini2026DUCsreview}. A typifying feature of circuits built out of these gates is that two-point correlation functions are restricted to 1D paths in space (the edge of the causal light cone, $v=1$) and hence can be efficiently computed \cite{bertini2019exactcorrfuncs}\footnote{Note that despite this analytical tractability, these models are highly (in some senses, maximally) quantum chaotic~\cite{bertini2018exactsff,bertini2021randomsffDUCs,claeys2020mvqcs}.}. In Ref.~\cite{kos2021ducpathint}, Kos \textit{et al.}\ identified families of dual-unitaries for which this behaviour is stable under small perturbations - the correlations are no longer confined to the light cone edge, but can still be computed (up to a small approximation error) by performing a discrete path integral - a path sum - over all the possible 1D paths.
\par 
Kos \textit{et al.}\ also found numerical hints that some perturbed models outside of their stable class were still captured by a path-integral formalism, but with ``thickened'' paths. This motivates a generalisation of their approach to a path-sum method with transfer matrices of some width $d>1$ (as opposed to single-qudit maps). DTOE can be considered such a generalisation, with the original Kos \textit{et al.}\ path-sum equivalent to $d=1$ DTOE.
\par
DTOE was tested using diameters in the range $d \in [1,10]$ on the kicked Ising model
\begin{equation}\label{eqn:pduc_kickedising}
        U_{KI}[J, B, h] = U_I[J, h] U_K[B] U_I[J, h],
\end{equation}
with
\begin{equation}
    U_I[J, h] = e^{-iJ Z\otimes Z - i\tfrac{h}{2}(Z \otimes 1 + 1 \otimes Z)},
\end{equation}
and
\begin{equation}
    U_K[B] = e^{-iB\left(X \otimes \mathds{1} + \mathds{1} \otimes X\right)},
\end{equation}
taken at $J = B = \pi/4 + \eta$ and $h = 1.2$, for $\eta \in [0, 0.5]$ to a time of $t=100$ circuit layers. The model is dual-unitary for $J = B = \pi/4$ i.e.\ at $\eta = 0$. The correlator $C_{XX}(x,t)$ was measured across the whole causal-light cone; for distances $t-x = x_- \in [0,21]$ from the causal light cone edge, the correlator could be exactly evaluated using a light-cone transfer matrix method. In Fig.~\ref{fig:pduc}, the $|C_{XX}(x,t)|_d$ data is shown for the largest value in this range, $x_- = 21$, at a value of $\eta = 0.24588$, accompanied by a plot of the relative errors $(|C_{XX}(x,t)| - |C_{XX}(x,t)_d|)/|C_{XX}(x,t)|$. For each $(\eta, d)$ pair, we also compute the median of the relative error distribution (over the $t\in[1,100]$ window) - this data is plotted in the right-hand panel of Fig.~\ref{fig:pduc}.
\par
In Ref.~\cite{kos2021ducpathint}, it was noted that the structure of the kicked Ising model can lead to the generation of quasiparticles with support greater than one that are relevant for the computation of the two-point functions. For $\eta = 0.24588$, the $|C_{XX}(x,t)|_d$ timeseries data shown in Fig.~\ref{fig:pduc} provides evidence for this, in that one can clearly see the low diameter data (for $d=1,2$ in particular) diverging from the exact data. Conversely, the higher diameters perform well, and we see consistent relative errors on the order of $10^{-2}-10^{-3}$ for the $d \geq 9$ data, even at this high (essentially non-perturbative) value of $\eta$. 
\par
Looking at the plot of the median relative errors against $\eta$, one can note that across the $\eta$ parameter space, DTOE seems to perform relatively well (median relative errors below $10^{-2}$, save for a small window around $\eta = 0.1$) across the $\eta$ parameter space. Though this is just a small snapshot, it suggests that DTOE can be effective for studying two-point functions in chaotic systems - or at the very least, in a large perturbative region around dual-unitary points - up to time-scales slightly beyond those attainable in experiments with current digital quantum computers.\footnote{Recent quantum simulations of chaotic circuits typically manage around 25 Floquet cycles~\cite{google2024dynamicsofmagnetization,fischerIBM2026dynamicalsimulationsofmbqc}, whereas here we reach $t/2 = 50$ cycles.}

\begin{figure*}
    \includegraphics[width=\textwidth]{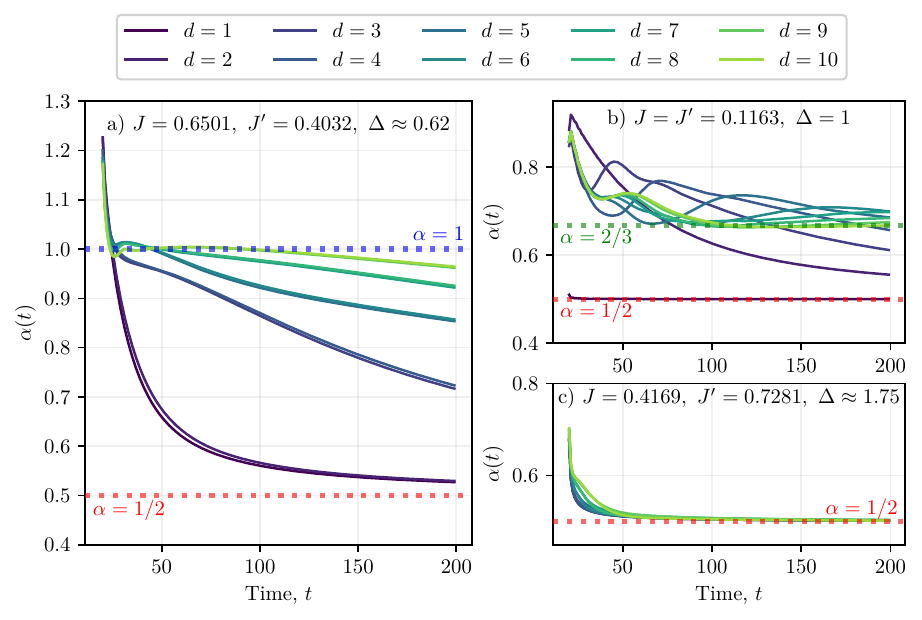}
    \caption{DTOE on the XXZ model. An example set of $d \in [1,10]$ time-series for the extracted scaling exponent $\alpha(t)$ in each of the three transport regimes are shown. All $\alpha_d(t)$ data is produced by fitting the tail (the window $[t-\delta, t]$ with $\delta=20$) of $\sigma_d(t)$ with a power-law fit and extracting the exponent. $\Delta < 1$. Ballistic transport ($\alpha = 1$) is expected. A value of $\alpha = 1.00 \pm 0.01$ was extracted from the $d=10$ data shown above. b) $\Delta = 1$. Superdiffusion ($\alpha = 2/3$) is expected. For $\delta \in [10,50]$) $d=10$, exponents of $\alpha \in [0.6663, 0.6684]$ were extracted from the $d=10$ data. c) $\Delta > 1$. Diffusion ($\alpha = 1/2$) is expected. DTOE reproduces this exponent accurately across all diameters.}\label{fig:xxz}
\end{figure*}
\begin{figure}
    \includegraphics[width=\linewidth]{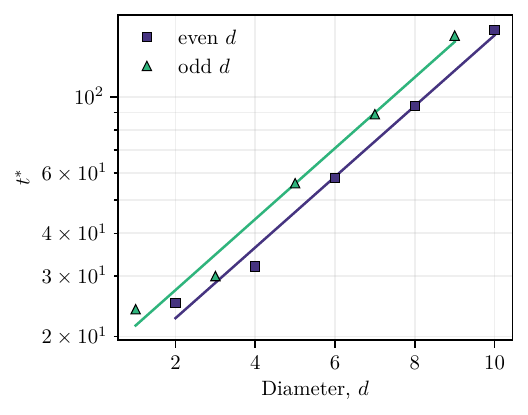}
    \caption{The breakdown timescale $t^*$ (Eqn.~\ref{eqn:tstar}) of DTOE in the ballistic ($\Delta < 1$) regime of the XXZ model, plotted against diameter $d$. The even and odd diameter data sets are fitted separately with linear fits. Both fits agree well with the data ($R^2 = 0.99$ for $d$ even, $R^2 = 0.98$ for $d$ odd), supporting an exponential relationship between $t^*$ and $d$.}\label{fig:timescale}
\end{figure}

\noindent \textit{Heisenberg XXZ}---DTOE was tested on circuits built out of gates of the form
\begin{equation}\label{eqn:heisenbergparam}
    U(J, J^{\prime}) = e^{-i\left[J(X \otimes X + Y \otimes Y) + J^{\prime}(Z \otimes Z)\right]}.
\end{equation}
This is the XXZ circuit model - for small $J$ and $J^{\prime}$ it corresponds concretely to a Trotterisation of the XXZ Hamiltonian. However, we make no attempt here to work strictly in the small-angle limit - the qualitative behaviour of the infinite-temperature spin dynamics should mirror that of the continuous case regardless \cite{ljubotina2019ballisticspintransport}. Specifically, the scaling behaviour of $C_{ZZ}(x,t)$ correlators ($ZZ$-correlators are the appropriate quantity here, since total $S_z$ magnetisation is conserved for all $J$,$J^{\prime}$) should fall in to three regimes, governed by the value of $\Delta = J^{\prime}/J$: 
\begin{itemize}
    \item $|\Delta| < 1$: ballistic, with dynamical exponent $\alpha = 1$;
    \item $|\Delta| = 1$ (XXX): superdiffusive, with $\alpha = 2/3$;
    \item $|\Delta| > 1$: diffusive, with $\alpha = 1/2$.
\end{itemize}
We probe this by computing the width of the $C_{ZZ}(x,t)$ profile at each time-step
\begin{equation}\label{eqn:width}
    \sigma(t) = \sqrt{\sum_x (x - \bar{x})^2 \, C_{ZZ}(x,t)},
\end{equation}
with
\begin{equation}
    \bar{x} = \sum_x x \, C_{ZZ}(x,t),
\end{equation}
which should scale as
\begin{equation}\label{eqn:alpha_def}
    \sigma(t) \sim t^{\alpha}.
\end{equation}
Fig.~\ref{fig:xxz} shows a sample of timeseries $\alpha(t)$ - one in each of the relevant transport regimes - extracted from the $\sigma(t)$ data. Each timeseries was extracted by fitting $\sigma(t)$ over the window $[t-\delta, t]$ with a power-law fit at each timestep and reading off the exponent. Multiple values of the parameter $\delta$ were tested, with $\delta = 20$ used for the plots in Fig.~\ref{fig:xxz}. An estimate for the true hydrodynamical scaling exponent $\alpha$ would then be extracted as the value of $\alpha(t)$ for the smallest $t$ such that $d\alpha(t)/dt < \epsilon$ i.e. once the timeseries had sufficiently plateaued. Again, multiple values of the parameter $\epsilon$ were tested, with $\epsilon = 10^{-4}$ used for the plots in Fig.~\ref{fig:xxz}.
\par
In the $\Delta < 1$ ballistic regime, a value of $\alpha = 1.00 \pm 1$ was extracted from the diameter $d=10$ data, in excellent agreement with the theoretically predicted value of $\alpha = 1$. At late-times, the data appears to drift down towards $\alpha = 1/2$, indicating diffusion. This indicates that the highest diameters used here are not enough to capture the full set of quasi-local charges that contribute to the transport \cite{ljubotina2019ballisticspintransport} - we are imposing a finite length-scale cut-off below the length of the largest strings that contribute to the transport, and lose the corrections produced by these strings necessary to generate $\alpha = 1$. The remaining strings left in the diameter-$d$ subspace below the cut-off appear to undergo random-walk dynamics (at least, this is the conclusion one draws from seeing $\alpha = 1/2$). Nonetheless, we seem to be able to simulate the system for long enough to obtain an accurate approximation to the correct $\alpha =1$ exponent. Additionally, we see evidence that the timescale on which DTOE accurately approximates the ballistic exponent scales exponentially with the diameter $d$. We compute the time $t*$, defined as
\begin{equation}\label{eqn:tstar}
    t^*(d) := \sup \{t : |\alpha_d(t) - 1| < \epsilon \;\},
\end{equation}
which is intended to serve as a rough indicator of when the diameter-$d$ DTOE simulation (in the ballistic regime) breaks down and begins to erroneously drift towards diffusion. The values of $t^*$ extracted from using $\epsilon = 0.02$ with the $\delta = 20$ $\alpha_d(t)$ data are plotted in Fig.\ref{fig:timescale}. The plot shows good agreement with an exponential fitting, suggesting $t^* = O(\exp(d))$, though the sample size is small ($n = 5$ points for the two different parity fits shown).
\par
At the isotropic point, $\Delta = 1$, superdiffusion - indicated by an exponent $\alpha = 2/3$ - is expected. The DTOE data for the point $J = J^{\prime} = 0.1163$ is shown in Fig.~\ref{fig:xxz}. Extracting $\alpha$ by fitting the late-time ($t^{\prime} \in [t-\delta,t]$ with $\delta \in [10,50]$) $d=10$ data, we get $\alpha \in [0.6663, 0.6684]$, in excellent agreement with the theoretically predicted value. The importance of large quasiparticles to the transport - as $\Delta \rightarrow 1$, quasiparticle strings (bound states of magnons) of all length scales are expected to contribute to the transport, and be critical to producing the superdiffusive exponent \cite{gopalakrishnan2019kinetictheoryXXZ} - makes this point trickier for DTOE in general, though for small enough values of the coupling strength we see (as shown in Fig.\ref{fig:xxz}) that the correct exponent can be captured.
\par
In the diffusive regime, $\Delta > 1$, diffusion is expected, indicated by an exponent of $\alpha = 1/2$. DTOE performs well here, reproducing this exponent accurately even at the lowest diameters. This is to be expected, as here the transport is dominated by short quasiparticle strings \cite{gopalakrishnan2019kinetictheoryXXZ}. We see that by a time of $t \approx 75$, all the $d \in [1,10]$ DTOE data has converged on $\alpha = 1/2$ in the $J = 0.4169, \; J^{\prime} = 0.7281, \; \Delta \approx 1.75$ example shown in Fig.\ref{fig:xxz}.

\noindent \textit{Summary}---In this work, we have introduced diameter-truncated operator evolution, a new method for simulating operator dynamics in local quantum systems. We have demonstrated its efficacy via computations of infinite-temperature two-point correlation functions in perturbed dual-unitary circuits and XXZ circuits, with hydrodynamical scaling exponents extracted in the latter example at late-times to a high degree of accuracy.
\par
The method is closely related to dissipation-assisted operator evolution (DAOE), differing in the parameter used for truncation (diameter versus Pauli weight). A further difference is that DAOE is cast as an MPO method, with the truncation of high-weight operators carried out by a dissipation superoperator that can be exactly represented as an MPO with a fixed bond dimension of $l_{*} + 1$, where $l_{*}$ is the cut-off weight.
\par
In principle, it should be possible to frame DTOE in the same way - the challenge is in representing the dissipation superoperator. We outline how this would be achieved in the Appendix, Sec.~\ref{sec:DTOEdissipator}. Implementing this would be desirable, as it would allow for extrapolation in the dissipation strength (as is done with DAOE), however we leave this to future work.
\par
We also note again that the approach outlined here would be particularly applicable, if an appropriate generalisation is possible, to higher lattice dimensions. In Ref.~\cite{nahum2022realtimecorrelatorsinchaoticmbs} it is argued that for \emph{generic} 2D circuits, the \emph{whole} of the lattice is in the easy-phase, where 2-point correlators are dominated by trajectories through operators with an $O(1)$-sized region of support. Again, the number of operators of a fixed space-time volume should scale fairly well - as $O(L^D)$ - when the number of spatial dimensions $D$ is low, presenting DTOE as a promising approach to computing 2-point correlators in this more general setting.

\noindent \textit{Acknowledgements}---THD acknowledges support from the EPSRC Centre for Doctoral Training in Delivering Quantum Technologies [Grant Number EP/S021582/1]. A.P. was funded by the European Research Council (ERC) under the European Union’s Horizon 2020 research and innovation programme (Grant Agreement No. 853368)

\bibliography{main}

\appendix

\section{Construction of the DTOE dissipator}\label{sec:DTOEdissipator}

As alluded to in the main-body of the text, DTOE is similar in spirit to the method of dissipation-assisted operator evolution, or DAOE \cite{rakovszky2018diffusive}. DAOE has been implemented as a method in which the initial state (a vectorisation of a local operator on the infinite-temperature state i.e. $\fket{B_0}$, as considered here) is represented as a matrix-product state, which is then evolved under the time-evolution of a local circuit according to the time-evolving block decimation algorithm \cite{vidal2003tebd}, but with a layer of dissipation applied after each layer of time-evolution. The dissipator is constructed so as to apply a penalty of $e^{-\gamma(l-l_*)}$ to each Pauli string $P^{(\vec{s})}$, where $l$ is the weight of the string (the number of non-trivial Pauli-terms), $l_*$ is a finite cut-off weight, and $\gamma \in \mathbb{R}_+$ is some dissipation strength, if and only if $l > l_*$. This dissipation can be achieved via an exact MPO representation with bond-dimension $l_* +1$ - see Ref.~ \cite{rakovszky2018diffusive} for an explicit construction.
\par
DTOE can be performed in a very similar way, but a new dissipator must be constructed (since we require a dissipator that dissipates based on the diameter of a string, rather than the weight). It turns out there is also an exact MPO representation for this, with a bond-dimension again one more than the value of the cut-off parameter, i.e. $d+1$. 
\par
\begin{figure*}
    \includegraphics[width=\textwidth]{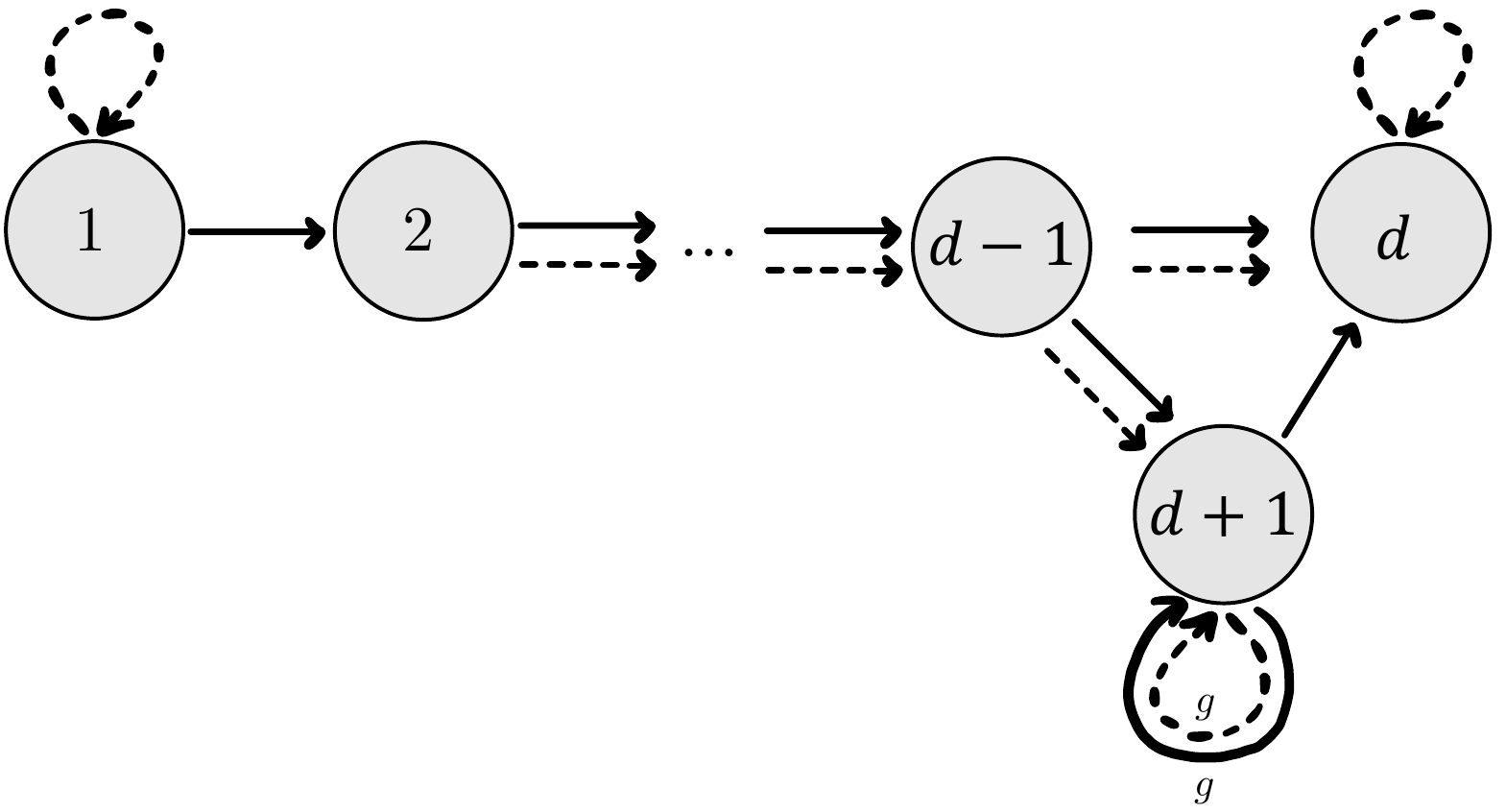}
    \caption{The automaton representing the construction of the DTOE dissipator $\mathcal{D}_{d,\gamma}$ as an MPO with bond dimension $d+1$, for $d\geq 3$. Transitions between the bond states generated by the presence of a $\mathds{1}$ term on the physical indices are shown by dashed arrows, and transitions generated by non-identity terms shown by full arrows.}\label{fig:automaton}
\end{figure*}

Following the notation used in DAOE, we have a local MPO tensor $W^{n,n^{\prime}}_{a,b}$ where the raised, physical indices are labelled by $n \in \mathbb{Z}_{q^2}$, i.e. labelling the $q^2$ generalised Pauli operators on the $q$-dimensional local site, and the lower indices correspond to bond states in $\{1,..d+1\}$. We define a dissipation strength $g = e^{-\gamma}$ for some $\gamma \in \mathbb{R}_+$. The local tensor is then constructed as follows:
\begin{equation}
    W^{\mathds{1},\mathds{1}}_{1,1} = 1;
\end{equation}
\begin{equation}
    W^{n\neq \mathds{1},n^{\prime}n\neq \mathds{1}}_{1,2} = \delta_{n,n^{\prime}};
\end{equation}
\begin{equation}
    W^{n,n^{\prime}}_{a,a+1} = \delta_{n,n^{\prime}} \quad \forall \; a \in \{2, d-1\};
\end{equation}
\begin{equation}
    W^{n,n^{\prime}}_{d-1,d+1} = \delta_{n,n^{\prime}};
\end{equation}
\begin{equation}
    W^{\mathds{1},\mathds{1}}_{d,d} = 1;
\end{equation}
\begin{equation}
    W^{n\neq \mathds{1},n^{\prime}\neq \mathds{1}}_{d+1,d} = \delta_{n,n^{\prime}};
\end{equation}
\begin{equation}
    W^{n,n^{\prime}}_{d+1,d+1} = g \delta_{n,n^{\prime}}.
\end{equation}
We construct the dissipator as
\begin{equation}
\begin{split}
    &\mathcal{D}_{\gamma,d}^{n_{-L+1},..,n_{L},n^{\prime}_{-L+1},...,n^{\prime}_{L}} = \\ & \; \; v^L_{a_{-L}}W^{n_{-L+1},n^{\prime}_{-L+1}}_{a_{-L},a_{-L+1}}W^{n_{-L+2},n^{\prime}_{-L+2}}_{a_{-L+1},a_{-L+2}}... \\
    & \; \; ...W^{n_{L-1},n^{\prime}_{L-1}}_{a_{L-2},a_{L-1}}W^{n_{L},n^{\prime}_{L}}_{a_{L-1},a_{L}}v^R_{a_{L}},
\end{split}
\end{equation}
where $v^L = \ket{1}$ and $v^R = \sum_{j=1}^d \ket{j}$. One can verify that for all $d \geq 3$, this produces the desired behaviour,
\begin{equation}
    \mathcal{D}_{d,\gamma}[\mathcal{S}] = \begin{cases}
        e^{-\gamma(d_{\mathcal{S}} - d)}\mathcal{S} &\; \textrm{if} \; d_{\mathcal{S}} > d \\
         \mathcal{S} &\; \textrm{otherwise}
    \end{cases},
\end{equation}
where we have used a shorthand $\mathcal{S} = P^{(\vec{s})}$ and $d_{\mathcal{S}} = \textrm{diam}(\mathcal{S})$. Without working through the algebra, however, one can gain some intuition for the construction by viewing it as a finite-state automaton, as shown in Fig. \ref{fig:automaton}. We start in the initial state $v_L = \ket{1}$. Moving along the lattice from left to right ($-L+1$ to $L$), we start counting by moving in to the $\ket{2}$ state on the bond space when we encounter our first non-identity term on the corresponding physical index. We then keep moving up the bond states (regardless of whether we see identity or non-identity terms on the physical indices), until we get to the penultimate site in our diameter-$d$ window, at which point we map from the state $\ket{d-1}$ into the $\ket{d}$ and $\ket{d+1}$ sites (again, regardless of what term we see on the physical index). The state $\ket{d}$ is to act as a state corresponding to the case where we have seen the last non-identity term on the lattice, where as the state $\ket{d+1}$ acts as an ``overflow" state that corresponds to the case where we have more non-identities to come. Whenever we see an identity term after this, we simply map back into both states, applying $g$ to the overflow state (so that it keeps track of the correct amount of dissipation). When we see a non-identity term, we kill off the $\ket{d}$ state (since the instance of having seen the last non-identity term it corresponded to has been invalidated), and map from the $\ket{d+1}$ state into the $\ket{d}$ state to start a new instance of having seen the last non-identity (but now with some amount of dissipation applied). At the end of the chain, we contract on to the $\ket{d}$ state (and all the states $\ket{i}$ with $i < d$, in case our first non-identity came close to the end of the chain and we haven't yet exited the diameter-$d$ window), as we want to record the state (with the correct number of dissipation terms $g$ applied) that corresponds to having seen our last non-identity (since we are at the end of the chain, and cannot possibly see anymore non-identity terms).
\par
\begin{figure*}
    \includegraphics[width=\textwidth]{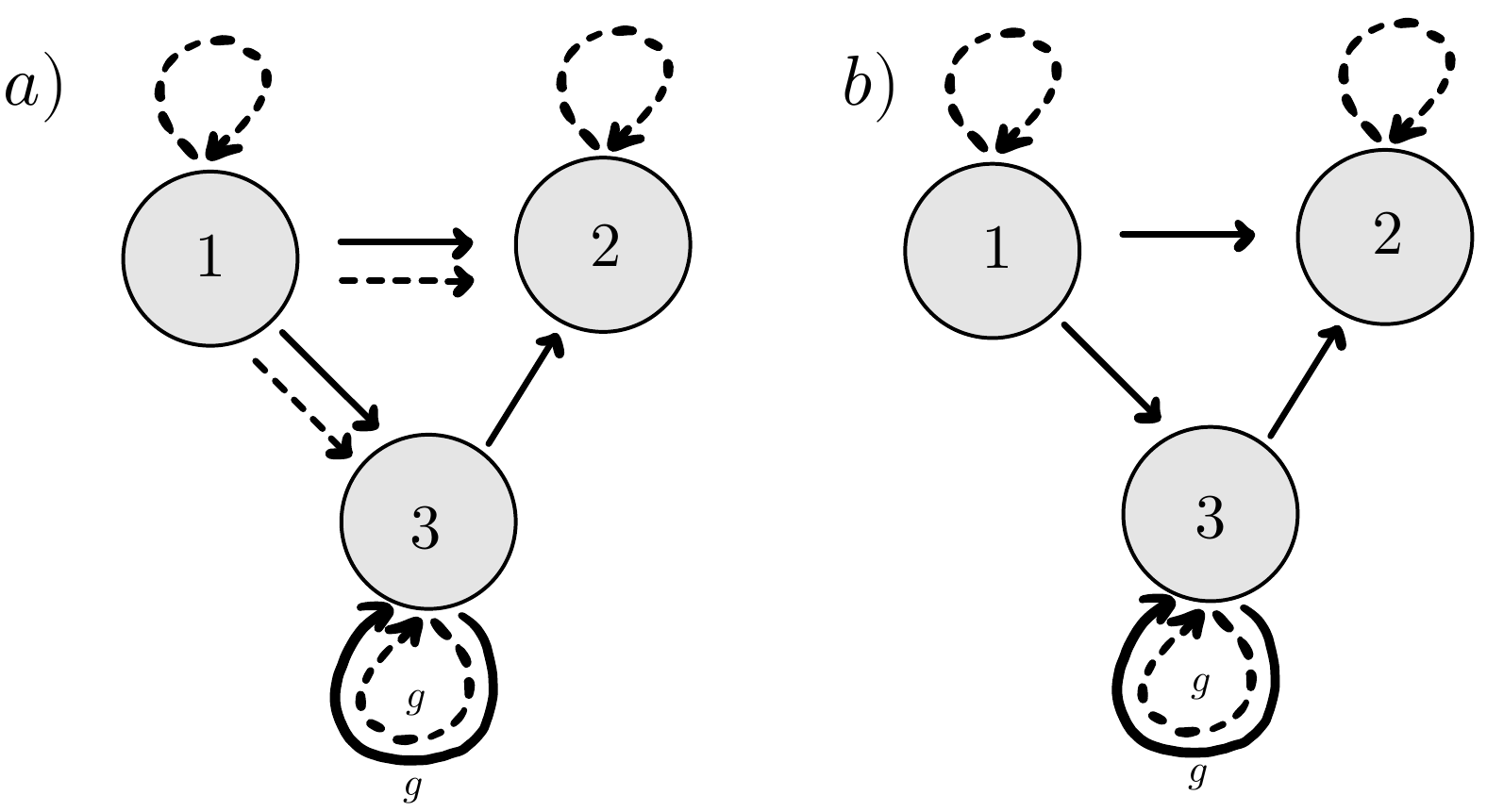}
    \caption{The automata for the edge cases $d=1$ and $d=2$. Again, transitions between the bond states generated by the presence of a $\mathds{1}$ term on the physical indices are shown by dashed arrows, and transitions generated by non-identity terms shown by full arrows. $a)$ The automaton for $d=1$. We could not find a bond-dimension $d+1 = 2$ representation for $\mathcal{D}_{1,\gamma}$, with the representation shown above requiring $\chi = d+2 = 3$ states. The state $\ket{d+1} = \ket{2}$ takes the role of the threshold (``we have seen the last non-identity") state, while $\ket{d+2} = \ket{3}$ takes the role of the overflow state. One would need to contract with $v^R = \sum_{j=1}^{d+1} \ket{j}$ on the right to account for this change in the structure of the bond space. $b)$ The automaton for the $d=2$ case. The structure is largely the same as in the $d\geq 3$ general case, but we only map into the $\ket{d+1}$ and $\ket{d}$ states upon seeing a non-identity term.}\label{fig:automatonedgecases}
\end{figure*}

There are edge cases for $d=1$ and $d=2$. The automata for these cases are shown in Fig.~\ref{fig:automatonedgecases}, from which the local MPO tensors can be derived.

\end{document}